\documentclass[12pt]{article}
\usepackage{a4wide}
\usepackage{graphicx}
\usepackage{amssymb}
\usepackage{amsmath}
\usepackage{slashed}
\usepackage{comment}
\usepackage{cite}
\usepackage{color}
\usepackage{multicol}
\usepackage{braket}
\usepackage{hyperref}
\usepackage[affil-it]{authblk}
\usepackage{ulem}
\newcommand{\be}{\begin{equation}}
\newcommand{\ee}{\end{equation}}
\newcommand{\ba}{\begin{eqnarray}}
\newcommand{\ea}{\end{eqnarray}}

\title{(sub)GeV Dark Matter in the $U(1)_X$ Higgs Portal Model}

\author{\small Amir Amiri$^1$\thanks{amir.amiri1308@gmail.com}, Basti\'an D\'iaz S\'aez$^2$\thanks{bastian.diaz.s@usach.cl}, Karim Ghorbani$^1$\thanks{karim1.ghorbani@gmail.com}}

\affil{\it $^1$Physics Department, Faculty of Sciences, Arak University, Arak 38156-8-8349, Iran}
\vspace{-0.6cm}
\affil{\it $^2$Departamento de Física, Universidad de Santiago de Chile, Casilla 307, Santiago, Chile}

\date{}

\begin{document}

\maketitle

\begin{abstract}
In this research we consider a $U(1)_X$ gauge boson acting as a dark matter candidate.
The vector dark matter (DM) gets mass when a complex singlet scalar breaks the gauge symmetry spontaneously, adding a second Higgs boson to the spectra. The dark matter candidates communicate with the SM particles via a scalar-Higgs portal. In this work, we concentrate on the masses of the vector dark matter and the scalar mediator below 10 GeV, aka \textit{light dark matter}. Although we assume thermal freeze-out for the vector DM using the zero-moment of the full Boltzmann equation to calculate the relic abundance, we explore the effects of the second-moment when the vector DM annihilates resonantly. As typically light DM is highly sensitive to CMB bounds, we focus on two thermal mechanisms which alleviate this bound: dark matter annihilation via forbidden channels and near a pole. Other bounds from colliders, thermalization conditions, beam-dump experiments, and astrophysical 
observations are imposed. Taking into account all the bounds including the direct detection upper limits, the viable space is achieved.   

\end{abstract}
\vfill

\vfill
{\footnotesize\noindent }


\vspace{4cm}

\tableofcontents

\section{Introduction}
The weakly interacting massive particles (WIMPs) as dark matter (DM) candidates come in various types of interactions with the normal matter \cite{STEIGMAN1985375,Arcadi:2017kky,Bergstrom:2000pn}. 
Recent improvements on the direct detection (DD) experiments have been able to exclude a large portion of the parameter space in minimal extensions of the Standard Model (SM). 
The minimal extensions are used to be applied in both simplified DM models and in effective field theory (EFT) DM models. 
The disadvantage of the latter approach is that the viable parameter space 
is shrunk by the constraints from the applicability criterion of the EFT models \cite{Arcadi:2021mag}. 

A generic question is that how small the WIMP mass might be when assuming a thermal mechanism  for the production of dark matter. 
We address this issue in the present work within a $U(1)_X$ vector DM model.
The model consists of a dark gauge boson acting as a DM candidate and a complex scalar which 
is responsible for the spontaneous breaking of the gauge symmetry and 
thereby giving mass to the dark matter candidate. DM particles can communicate with the normal matter via scalar-Higgs portal. The abelian vector DM model is studied earlier \cite{Lebedev:2011iq,Baek:2012se,Farzan:2012hh,Glaus:2019itb,Arcadi:2016qoz,Arcadi:2020jqf,Ghorbani:2021yiw,Duch:2015jta,YaserAyazi:2022tbn, Foguel:2022unm}. 
In all the previous works, points in the parameter space with quite small singlet scalar mass and at the same time very small mixing angle are not considered. In \cite{Yang:2022zlh} the same scenario is studied including number changing interactions.

In this work, we study the parameter space of the vector DM assuming thermal freeze-out, respecting the observed relic density, and taking into account the following. First, we calculate the relic abundance via the standard procedure making use of MicrOMEGAS code (these results are used in most of the paper). Secondly, as resonantly DM annihilation via freeze-out tends to decrease the couplings which fulfill the correct relic abundance, this makes a strongly suppressed momentum transfer rate $\gamma(T)$ at the freeze-out epoch, eventually producing an \textit{early kinetic decoupling} of the DM particle\cite{Binder:2017rgn}. In this case, we obtain the evolution of the system by solving the second-moment of the full Boltzmann equation. On the other hand, upper bounds on the annihilation cross section from Cosmic Microwave Background (CMB) have shown to be particularly important for light DM, therefore we include those constraints in this work. To address this issue, one has to consider the forbidden and near pole annihilation channels carefully. 
There are also various constraints from collider searches, thermalization, beam-dump experiments, and astrophysical observations, which are incorporated in this work. 

The paper has the following structure. The DM model is described in the next section along with theoretical bounds on the couplings from vacuum stability. In section \ref{Experimental-bounds}  we show the relevant constraints from collider and beam-dump. Focusing on the low mass DM, we present the relevant Feynman diagrams for the DM annihilation and study the behavior of the DM relic density in section \ref{Rdensity} based on both standard approach or zero-moment of the full Boltzmann equation, and the second-moment of the latter. Bounds from CMB are studied in \ref{CMBbounds}, restricting notably the full available parameter space. In section \ref{Secresults} we resume the viable parameter space found after imposing all the constraints, and the conclusions are provided in section \ref{con}. In the Appendix we provide the DM annihilation cross sections, temperature dependent cross sections, and thermalization conditions.

\section{Model}
\label{model}
We consider a model as a minimal extension to the SM with a new $U(1)_X$ gauge symmetry introducing a complex scalar field, $S$, in the following Lagrangian 
\begin{equation}
{\cal L}_{\text{DM}}   = ({\cal D}_\mu S)({\cal D}^\mu S)^* 
 - \frac{1}{4} X^{\mu\nu} X_{\mu\nu} - {\cal V } (S,H)    \,,
\end{equation}
where the covariant derivative is ${\cal D}_\mu = \partial_\mu - i \alpha_X q_X X_\mu$, in which $\alpha_X$ is the gauge coupling and $q_X$ is the $U(1)_X$ charge of the corresponding field which is taken unity. The strength field tensor for the gauge field $X_\mu$ is given by $X_{\mu \nu} = \partial_{\mu} X_{\nu} -\partial_{\nu} X_{\mu}$.   
The scalar potential entailing the Higgs portal is    
\begin{equation}
 {\cal V} (S, H) = \mu_H |H|^2 + \lambda_H |H|^4 + m^2 |S|^2 + \lambda_S |S|^4 + \lambda_{HS} |S|^2 |H|^2  \,.
\end{equation}

The Lagrangian enjoys an additional {\it charge conjugation} symmetry
\begin{equation}
 X^\mu  \to -X^{\mu},    ~~~~ S \to S^* \,.
\end{equation}
As discussed in \cite{Ma:2017ucp} the kinetic mixing term will break the dark charge conjugation symmetry, so if we choose the symmetry to be exact then the mixing term will be forbidden. 
In case we relax that symmetry and consider the gauge invariant kinetic mixing term, then the lifetime of dark vector boson has to be at the same order of the age of universe, to be a good candidate of dark matter. For the dark vector boson of mass $\sim$ GeV, it requires for the kinetic mixing angle to be less than $10^{-21}$.


When the complex scalar gets a nonzero expectation value this breaks the gauge symmetry spontaneously, after which the remaining residual symmetry for the vector gauge boson 
is $X^\mu  \to -X^{\mu}$. The parametrization of the complex scalar field around its vacuum is 
\begin{equation}
 S = \left( \frac{v_s + \phi_1}{2}\right) e^{-i \phi_2 /v_{s}} \,,
\end{equation}
here $\phi_1$ and $\phi_2$ are real scalar fields. It is possible to gauge away
the Goldstone field $\phi_2$ and rendering the massless gauge boson massive, such that $m_X = v_{s} \alpha_X$. On the other hand, the Higgs doublet takes a vacuum expectation value, $v_h$, and in unitary gauge we write down the Higgs field as $H = (0~~v_h + h_1)^T/ \sqrt{2}$. Plugging the Higgs doublet and 
singlet scalar in the Lagrangian and after some simplification we arrive at the following interaction terms, 

\begin{equation}
\begin{split}
 {\cal L}_{int} = & \alpha^2_X v_s X_\mu X^\mu \phi_1 
     +\frac{1}{2} \alpha^2_X X_\mu X^\mu \phi_1^2 
     - \frac{1}{4} \lambda_S \phi_1^4 - \lambda_S v_s \phi_1^3 
     - \frac{1}{4} \lambda_H h_1^4 - \lambda_H v_h h_1^3  \\
   &\quad  
     - \frac{1}{2} \lambda_{HS} v_h h_1 \phi_1^2 
     - \frac{1}{2} \lambda_{HS} v_s h_1^2 \phi_1 
     - \frac{1}{4} \lambda_{HS} h_1^2 \phi_1^2  \,.
\end{split}
\end{equation}
The scalar mass matrix is also obtained as 
\begin{equation}
 {\cal M}^2 = \Big( \begin{matrix}
              2\lambda_S v_s^2 & \lambda_{HS} v_h v_s \\
              \lambda_{HS} v_h v_s  & 2 \lambda_H v_h^2 
             \end{matrix} \Big) \,.
\end{equation}
A scalar field transformation is needed in order to diagonalize the mass matrix. 
Therefore, the physical or mass eigenstates, $h$ and $\phi$, are defined in terms of 
the mixing angle $\epsilon$, 
\begin{equation}
 h = h_1~\cos(\epsilon) + \phi_1~ \sin(\epsilon),  ~~~~  \phi  = -h_1 \sin(\epsilon) + \phi_1 \cos(\epsilon) \,,  
\end{equation}
where the mixing angle is given by the relation,
\begin{equation}
\tan(2\epsilon) = \frac{2v_h v_s}{m_h^2-m_{\phi}^2 }  \lambda_{HS} \,,
\end{equation}
with $m_h$ and $m_\phi$ the physical masses of the SM Higgs and the singlet scalar, respectively.
The quartic couplings are obtained in terms of the mixing angle and the physical scalar masses

\begin{equation}
 \begin{split}
  \lambda_H =& \frac{1}{2v_h^2} (m_h^2 \cos^2 \epsilon + m_{\phi}^2 \sin^2  \epsilon), \\
 \lambda_S =& \frac{1}{2 v_s^2} (m_h^2 \sin^2 \epsilon + m_{\phi}^2 \cos^2  \epsilon), \\
 \lambda_{HS} =& \frac{\sin 2\epsilon}{4v_h v_s} (m_{\phi}^2- m_h^2) \,. 
 \end{split}
\end{equation}
We will impose the conditions by which the Higgs potential remains bounded from below. These conditions read 
\begin{equation}
\label{BFM}
\lambda_H > 0,~~ \lambda_S > 0,~~ \lambda_{HS} > -2 \sqrt{\lambda_H \lambda_S}  \,.
\end{equation}
Also, satisfying the relations in Eq.~\ref{BFM} warrants the electroweak vacuum to be a global minimum \cite{Baek:2012se}.
The free parameters in this model are $m_X, m_\phi, \epsilon,$ and $\alpha_X$.

\section{Collider and Beam-dump Bounds}
\label{Experimental-bounds}
In this section we provide bounds from various experiments when 
dark particles are engaged in the decay or production at collider or beam-dump experiments.

\subsection{Invisible Higgs Decay}
\label{Inv}
In the mass range of interest in this work there are two channels for the SM Higgs invisible decay: $\text{Higgs} \to \phi \phi$ and $\text{Higgs} \to X X$.
The invisible Higgs decay width for decay a pair of dark matter is
\begin{equation}
 \Gamma_{h \to XX} = \frac{\alpha_X^2 m_h^3 \sin^2 \epsilon}{16\pi m_X^2} \sqrt{1-4(m_X/m_h)^2}  [1-4(m_X/m_h)^2+12(m_X/m_h)^4].
\end{equation}
The invisible Higgs decay width for the process $H \to \phi \phi$ is 
\begin{equation}
 \Gamma_{h \to \phi \phi} = \frac{c^2}{8\pi m_h} \sqrt{1-4(m_\phi/m_h)^2} \,,
\end{equation}
where the couplings $c$ is given by the relation below
\begin{equation}
 \begin{split}
   c =& \lambda_{HS} v_h \cos \epsilon  
     + 2 \lambda_{HS} v_s \sin \epsilon
     -3 v_s \lambda_{HS} \sin^3(\epsilon) + 6 \lambda_H v_h \sin^2(\epsilon) \cos(\epsilon) \\
  &\quad    -3  \lambda_{HS} v_h \sin^2(\epsilon) \cos(\epsilon) 
     -6  v_s \lambda_S  \cos^2(\epsilon) \sin(\epsilon) \,.
 \end{split}
\end{equation}
The measured Higgs mass is $m_h \sim 125.25$ GeV and the total Higgs decay width reads $\Gamma_{\text{Higgs}} = 3.2^{+2.8}_{-2.2}$ MeV \cite{ParticleDataGroup:2020ssz}.
The experimental upper bound on the branching ratio of the invisible Higgs decay  
at $95\% $ confidence level is Br($h\to$ invisible) $\lesssim 0.19$ 
\cite{CMS:2018yfx}.  
Thus, the following constraint is applied when the parameter space is scanned
\begin{equation}
 \frac{\Gamma_{h \to XX} + \Gamma_{h \to \phi \phi}}{\Gamma_{\text{Higgs}}} \lesssim 0.19 \,.
\end{equation}

\subsection{Singlet Scalar Decay}
\label{Sdecay}
Since the singlet scalar mass varies in the range $(\sim 0.1-10)$ GeV, when $m_\phi > 2 m_f$ it can decay to the SM fermions by the following decay width  
\begin{equation}\label{phiff}
 \Gamma (\phi \to f^+ f^- ) = \frac{N_c m_\phi m_f^2 \sin^2(\epsilon)}{8\pi v_h^2} 
    \Big(1- \frac{4m_f^2}{m_\phi^2} \Big)^{3/2} \,, 
\end{equation}
The total decay width is $\Gamma = \sum_{f} \Gamma (\phi \to f^+ f^- )$, and the corresponding life time of the scalar is $\tau = \Gamma^{-1}$.
Note that the decay $\phi \to XX$ is kinematically open when $m_\phi > 2m_X$.
The decay width is obtained as
\begin{equation}
 \Gamma (\phi \to XX) = \frac{\alpha_X^2 m_\phi^3 \cos^2 \epsilon}{16\pi m_X^2} \sqrt{1-4(m_X/m_\phi)^2}  [1-4(m_X/m_\phi)^2+12(m_X/m_\phi)^4] \,.
\end{equation}
The hadronic decay of the singlet scalar with final states, $\pi \pi$, $K K$,..., 
is also important here and has to be taken into account. 
The big bang nucleosynthesis (BBN) predicts the light element abundances in agreement with the observation. This puts upper limit on the singlet scalar life time which 
varies between $\sim 0.01$ sec to 1 sec, in the scalar mass range $\sim 0.1-4$ GeV \cite{Fradette:2017sdd}. This upper bound on the scalar life time will be converted to lower bounds on the mixing angle.

\subsection{Mediator Production in $e^+ e^-$ and LHC Colliders} 
Direct production of the dark mediator, $\phi$, mixing with the SM Higgs
is possible at colliders like LEP through the process $e^+ e^- \to Z \to \phi Z^*$ \cite{L3:1996ome}. Since the rate of the process is proportional to $\sin^2 \epsilon$, then strong constraint on the mixing angle, $\epsilon$, is expected. 
However, for mediator mass smaller than 10 GeV, it is found that regions 
with $\sin \epsilon \gtrsim 0.1$ are excluded \cite{Winkler:2018qyg}.

On the other hand, CMS and LHCb in search for a new scalar mediator 
put constraints on $\sigma_{pp-\phi} \times \text{Br}(\phi \to \mu^+ \mu^-)$ 
at $\sqrt{s} = 7$ and 8 TeV, respectively \cite{CMS:2012fgd,LHCb:2018cjc}.
It turns out that the LHC bounds are stronger than that of LEP for scalar mass smaller than about 5 GeV. 

\subsection{Rare Decays}
The relevant rare decays are $\Upsilon$ decay and flavor changing decay of $B$ and $K$ mesons. In these processes the singlet scalar can act as a mediator.
The most important decay channel of $\Upsilon$ studied by BaBar is $\Upsilon \to \gamma \phi^*$, where $\phi$ decays hadronically to jets \cite{BaBar:2011kau}.  
LHCb provides us constraints on two $B$ meson decays, $B^0 \to K^{0*} \mu^+ \mu^-$ and $B^+ \to K^+ + \mu^+ \mu^-$, where in both decays the scalar mediator $\phi$ is decaying to a dilepton \cite{LHCb:2015nkv,LHCb:2016awg}. 
When the scalar mass is below a few hundred of MeV, rare kaon decays become important. Constraints on the mixing angle in terms of the scalar mass can be obtained from the decay $K_L \to \pi^0 \mu^+ \mu^-$ performed by KTeV experiment \cite{KTEV:2000ngj} and the search for the decay $K^+ \to \pi^+ \bar \nu \nu$ by E949 experiment \cite{BNL-E949:2009dza}. 

\subsection{Beam Dump Experiments}
We apply bounds from three beam dump experiments at CERN SPS using a 400 GeV proton beam, see \cite{Winkler:2018qyg} and references therein. 
1) CHARM experiments being sensitive to leptonic final states was 
operating in 1980s. 2) NA62 in dump mode
which is an upcoming run and as CHARM is sensitive to higher scalar mass. 3) SHiP is a planned experiment which will cover a large region 
of the parameter space not accessible in all earlier experiments.

\section{Relic Density from Thermal Freeze-out}\label{Rdensity}
\subsection{Zero-moment of the Boltzmann equation}
\begin{figure}
\begin{center}
\includegraphics[width=.7\textwidth,angle =0]{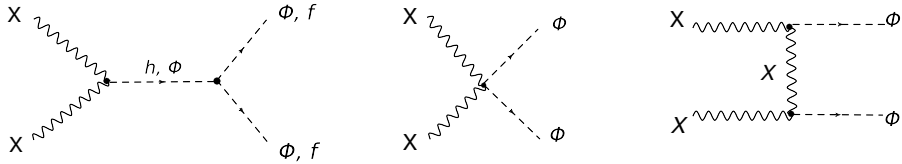}
\caption{The relevant Feynman diagrams for the vector dark matter annihilation with mass smaller 
than $\sim 10$ GeV.
The diagram in the left is a $s$-channel annihilation process and the process shown in the right side takes place in both $t$- and $u$-channel.}
\label{FeynAnni}
\end{center}
\end{figure}

\begin{figure}[h!]
\centering
\includegraphics[width=0.42\textwidth]{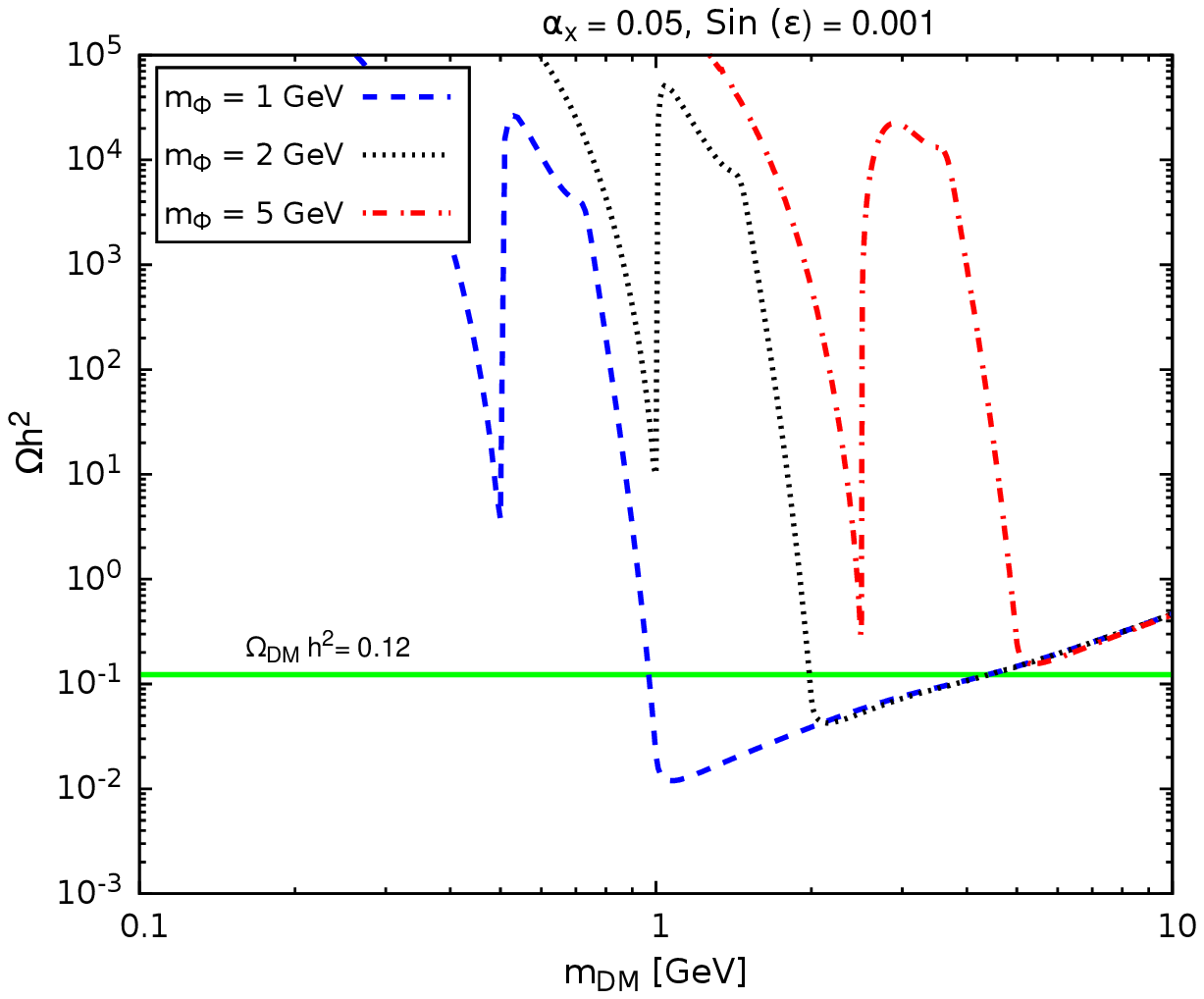}
\includegraphics[width=0.42\textwidth]{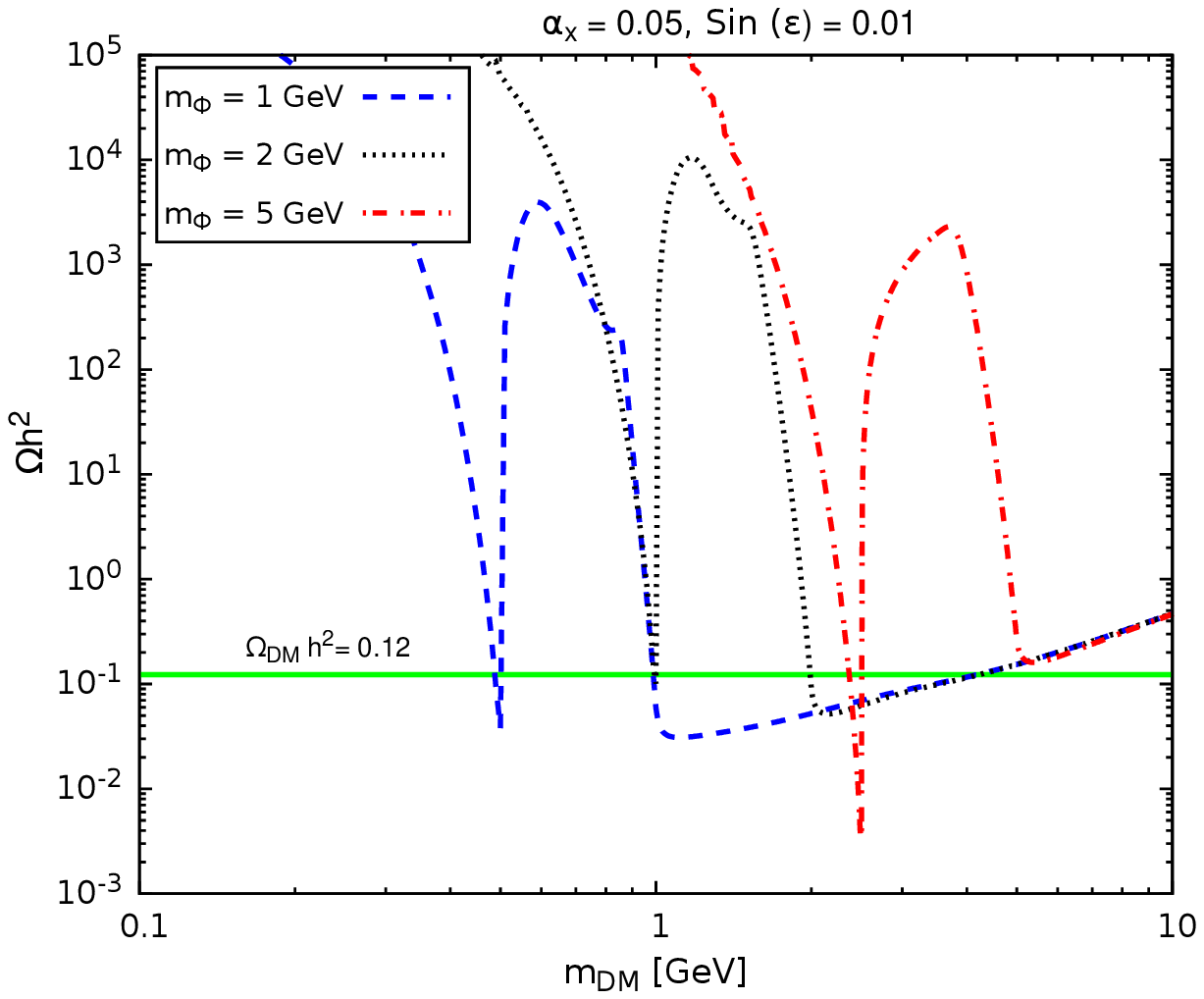}
\includegraphics[width=0.34\textwidth, angle=-90]{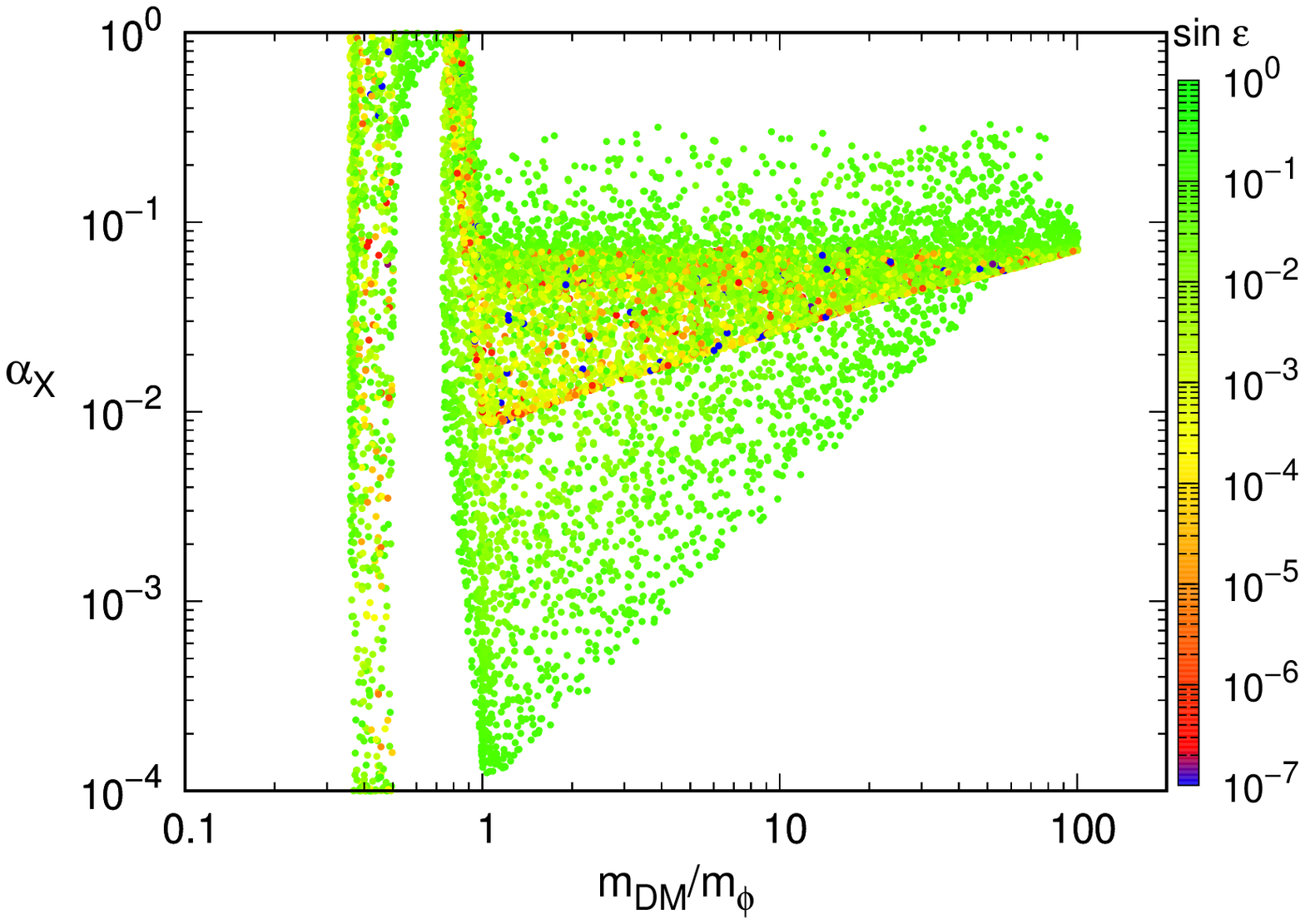}
\caption{{(above) Shown are relic density as a function of DM mass for the gauge coupling $\alpha_X = 0.05$. 
In the left panel we take the mixing angle such that $\sin \epsilon = 0.001$ 
and in the right panel we fix it as $\sin \epsilon = 0.01$. The observed DM relic density is shown as a horizontal line at $\Omega h^2 \sim 0.12$. (below) Full random scan with each point fulfilling the measured relic abundance $\Omega h^2 \sim 0.12$.}}
\label{relicdensity}
\end{figure}

Assuming DM is in thermal equilibrium in the early times, dark matter production 
is then happening at temperatures smaller than 
the DM mass when it goes out of equilibrium, the so-called freeze-out mechanism of dark matter particles. 
The freeze-out temperature, $T_f$, depends on the DM annihilation cross section and DM mass. 
The Boltzmann equation describes how the DM number density, $n_X$, evolves in an 
expanding Universe, 
\begin{equation}\label{sBE}
 \dot n_{X}  +3 {\cal H} n_{X} = - \langle \sigma_{\text{ann}}v_{\text{rel}} \rangle [n^{2}_{X}-(n_{X,e})^2 ] \,,
 \end{equation}
where ${\cal H}$ is the Hubble parameter, $n_{X,e}(T) = \frac{m_X^2}{2\pi^2}TK_2\left(\frac{m_X}{T}\right)$ is the equilibrium number density of the DM particle, and the 
thermal average annihilation cross section times the relative velocity is given by
\begin{equation}\label{sigv}
 \langle \sigma_{\text{ann}} v_{\text{rel}} \rangle = \frac{1}{8 m_{X}^4TK^{2}_{2}(\frac{m_{X}}{T})}
\int^{\infty}_{4m^{2}_{X}} ds~(s-4m^{2}_{X})\sqrt{s}~K_{1} \left(\frac{\sqrt{s}}{T} \right)~\sigma_{\text{ann}}(s)\,. 
\end{equation}
In the present work DM with mass smaller than $\sim 10$ GeV annihilates through three different ways diagrammatically as depicted in Fig.~\ref{FeynAnni}: 
1) A $s$-channel annihilation process into singlet scalars and the SM fermions 
as $XX \to \phi \phi, f \bar f$ with mediators being the SM Higgs or the singlet scalar. 
2) A contact interaction of DM with the singlet scalars. 
3) DM annihilation to singlet scalars via $t$- and $u$-channels, with the vector DM acting as the mediator.   
It is worth mentioning that the lower limit  for a thermal WIMP with $2 \to 2$ s-wave annihilation to visible final states is found to be about 20 GeV \cite{Leane:2018kjk}.
However, since in the present work we consider the forbidden channels and the resonance effect this lower bound is relaxed. 
This model is first implemented in the package LanHEP \cite{Semenov:2008jy} 
in order to obtain 
the possible vertices and then the code micrOMEGAs \cite{Belanger:2013oya} is applied to compute the relic density numerically. First, it is interesting to 
see the variation of the relic density as a function of DM mass for three different singlet scalar 
masses $m_\phi = 1, 2, 5$ GeV. To do that we fix the gauge coupling at a reasonable 
value $\alpha_X = 0.05$. The results are shown for two distinct mixing angles 
$\sin \epsilon = 0.001$ and $\sin \epsilon = 0.01$ in Fig.~\ref{relicdensity}. 

From these results we notice that, independent of the parameter selection, the relic density behavior results in similar patterns, and the correct relic abundance is obtained for certain specific DM mass values: (i) $m_{X} \sim m_\phi/2$, (ii) $m_X \lesssim m_\phi$, and (iii) $m_X > m_\phi$. In the first case, it is clear that DM annihilates \textit{resonantly} via $\phi$, then the latter decaying into SM states. In this way, small $\epsilon$ makes the resonant $s$-channel annihilation less effective, then resulting in a higher relic abundance (plot in the upper left of Fig.~\ref{relicdensity}). In the second case, $XX \to \phi \phi$ annihilation becomes relevant. Notice that the correct relic abundance can be obtained even when $m_X < m_\phi$ due to the thermal tail with high velocity $X$'s. This case has been dubbed as \textit{Forbidden dark matter}\cite{DAgnolo:2015ujb}\footnote{Notice that 3-body processes are subleading in comparison to $2\rightarrow 2$ forbidden annihilations \cite{Delgado:2016umt}.} In the third case, standard DM annihilation into lighter $\phi$ states. Later, we will see that only the first two cases are able to fulfill strong CMB bounds for light DM.

Before moving on, we want to complement the previous analysis with the results obtained in Fig.~\ref{relicdensity}(below), where we show a random scan in the full parameter space, with each point matching the correct relic abundance, and the color indicating the value of $\epsilon$. The information obtained in this plot reflects from another perspective what we have already obtained in the plots in the first raw of Fig.~\ref{relicdensity}: three regions with the correct relic abundance. Notice that, however, the mass window for the resonance and forbidden region is quite limited, whereas the case of $m_X > m_\phi$ spans much more masses.

\subsection{Second moment of the Boltzmann equation}
In this subsection we explore the deviations in the relic abundance and temperature of the dark sector when the assumption of kinetic equilibrium DM during chemical decoupling is not longer valid, i.e. the phase-space distribution $f_X(\bold{p})$ becomes $f_X \neq A(T)f_{X,eq}$, with $A(T) = 1$ in full equilibrum for $T \sim m_X/x_f$, with $x_f \sim 20$ (as it is usual, we define $x \equiv m_X/T$, with $T$ the SM photon temperature). It is well known that the system of Boltzmann equations for the zero and second moment (cBE) of the phase space distribution of the DM particle gives a reliable track of the DM yield $Y = n_X/s$, where $s(T) = (2\pi^2/45) g_{*s} T^3$ is the entropy density and $g_{*s}(T)$ the effective degrees of freedom in entropy at temperature $T$ \cite{Husdal:2016haj}, and the dark sector temperature quantified by the dimensionless parameter $y$ given by \cite{Binder:2017rgn}
\begin{eqnarray}
 y \equiv \frac{m_X}{3s^{2/3}}\Braket{\frac{\bold{p}^2}{E}} = \frac{m_X}{3s^{2/3}} \frac{g_\chi}{n_X}\int \frac{d^3p}{(2\pi)^3}\frac{\bold{p}^2}{E}f_X(\bold{p}).
\end{eqnarray} 
Equivalently, the dark sector temperature $T_X$ is given by $T_X=ys^{2/3}/m_X$. Furthermore, $Y_{e}(T) \equiv n_{e}/s$ as the equilibrium yield of the DM, and $y_e(T) \equiv m_X T/s^{2/3}$ tracking the photon temperature, the cBE to be solved is
\begin{eqnarray}\label{cBE}
 \frac{Y'}{Y} &=& \frac{sY}{x\tilde{H}} \left[\frac{Y_e^2}{Y^2}\braket{\sigma v} - \braket{\sigma v}_{neq}\right],  \\
 \frac{y'}{y} &=& \frac{\gamma(T)}{x\tilde{H}} \left[\frac{y_e}{y} - 1\right] + \frac{sY}{x\tilde{H}}\left[\braket{\sigma v}_{neq} - \braket{\sigma v}_{2,neq}\right] + \frac{sY}{x\tilde{H}}\frac{Y_e^2}{Y^2}\left[\frac{y_e}{y}\braket{\sigma v}_2 - \braket{\sigma v}\right] + \frac{H}{x\tilde{H}}\frac{\braket{p^4/E^3}_{neq}}{3T_X}, \nonumber
\end{eqnarray}
with $' \equiv \frac{d}{dx}$, and $\tilde{H} = H/[1 + \tilde{g}(T)]$, where $\tilde{g} \equiv \frac{T}{3g_{*s}}\frac{dg_{*s}}{dT}$. The initial conditions are $Y(T = m_X) = Y_{e}(m_X)$ and $y(T = m_X) = y_{e}(m_X)$. The expression for $\braket{\sigma v}$ and $\braket{\sigma v}_2$ are given in eq.\ref{sigv} and eq.\ref{sigv2}, respectively, and we take the ansatz from \cite{Binder:2017rgn} for $\braket{\sigma v}_{neq}$ and $\braket{\sigma v}_{2,neq}$. As we are interested in the resonant annihilation cross section $X+X\rightarrow \phi\rightarrow$ SM+SM, we take the narrow-width approximation (NWA) for each case, i,e. 
\begin{eqnarray}\label{nwa}
 \frac{1}{(s - m_{\phi})^2 + m_\phi^2 \Gamma_\phi^2} \approx \frac{\pi}{m_\phi \Gamma_\phi}\delta(s - m_\phi),
\end{eqnarray} 
where $\Gamma_\phi$ is the total width of the light scalar, including its decay into a pair of vector DM particles\footnote{For the parameter space that we are interested in, we have used CalcHEP code to check out that $\Gamma_\phi \lesssim m_\phi$, then supporting the approximation \ref{nwa}.}. The NWA approximation collapses the integral over the variable s into the average annihilation cross section, which is an algebraic expression. For the cross sections out-of-the-equilibrium and $\braket{p^4/E^3}_{neq}$, we take the ansatz given in \cite{Binder:2017rgn}, with each quantity evaluated at $T_X = ys^{2/3}/m_X$. The momentum exchange rate $\gamma(T)$ is given by
\begin{eqnarray}
 \gamma(T) = \frac{1}{48\pi^3 g_\chi m_X^3}\int_{0}^\infty dw \frac{1}{e^{w/T} + 1}\partial_w \left(k^4 \braket{|\mathcal{M}|^2}_t \right),
\end{eqnarray} 
where the integration takes the energy $w$ of the corresponding SM fermion in the plasma, and $\braket{|\mathcal{M}|^2}_t$ is the average of the invariant amplitude over the transferred momentum,
\begin{eqnarray}
 \braket{|\mathcal{M}|^2}_t \equiv \frac{1}{8k^4}\int_{-4k_\text{cm}^2}^0 dt (-t)|\mathcal{M}|^2,
\end{eqnarray}
with $k_{cm}^2 = \left(s - (m_X - m_f)^2\right) \left(s - (m_X + m_f)^2\right)/(4s)$, and the Mandelstam variable $s$ is evaluated at $s = m_\chi^2 + 2wm_\chi + m_f^2$\footnote{Since the physics is evaluated at $T \lesssim T_{f.o.}$, the DM $X_\mu$ is considered to be in rest.}. We have calculated the square of the invariant amplitude of the elastic process $X_\mu + f \leftrightarrow X_\mu + f$, with $f$ being a SM fermion from the thermal plasma that has not undergone annihilation, using CalcHEP, with the result
\begin{eqnarray}
 |\mathcal{M}|^2 = \frac{e^2 \alpha_X^2m_f^2 \sin^2\epsilon(-1 + \sin^2\epsilon)(4m_f^2 - t)(12m_X^4 - 4m_X^2 t + t^2)}{12(-1 + c_w^2)m_X^2 m_W^2(t - m_{\phi}^2)^2}.
\end{eqnarray}
\begin{figure}[h!]
\centering
\includegraphics[width=0.9\textwidth]{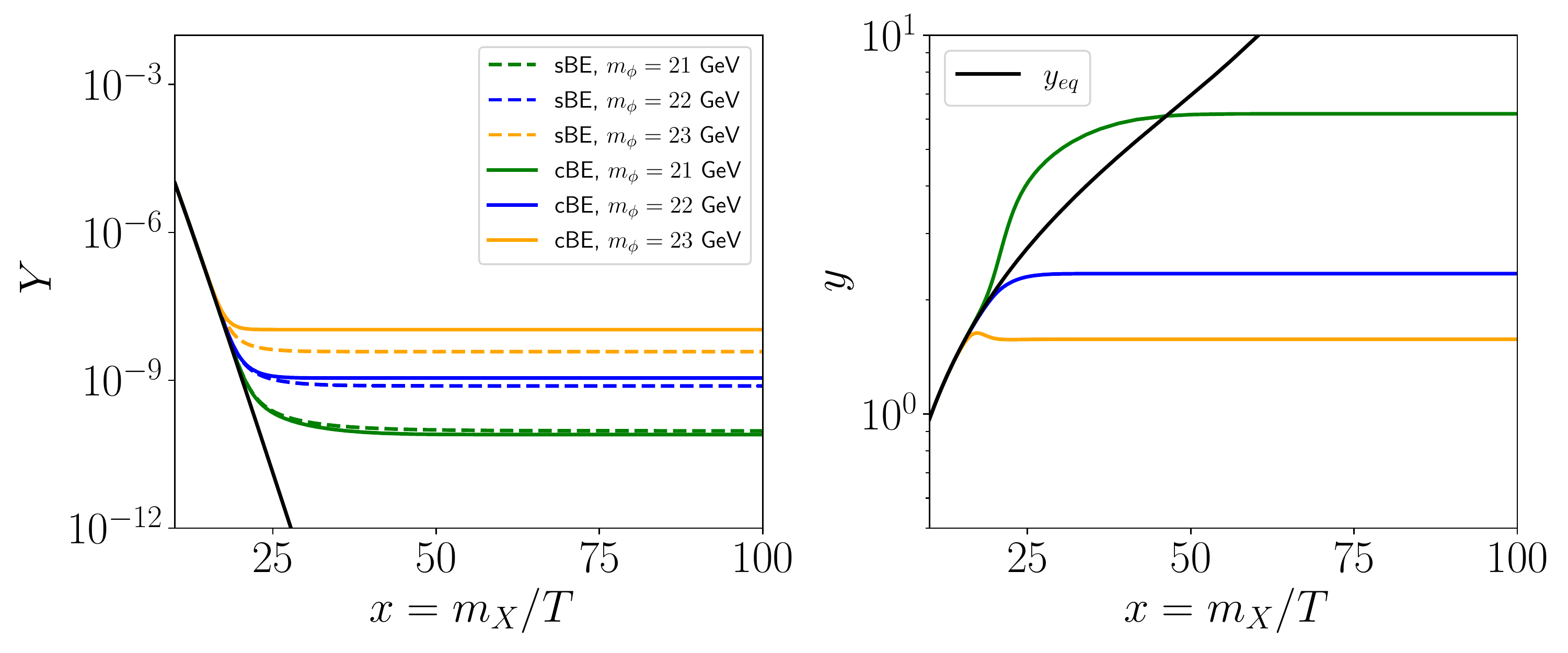}
\caption{Evolution of $Y$ and $y$ as a function of the photon temperature using the standard Boltzmann equation (sBE) and the coupled Boltzmann equation system (cBE) for $m_X = 10$ GeV, $\alpha_X = 0.5$, $\epsilon = 0.008$, and  the values of $m_{\phi}$ indicated in the legend of the plot in the left. The black lines in each plot shows the corresponding equilibrium values.}
\label{cBEfig}
\end{figure}
In the previous result we have integrated out the Higgs exchange in the elastic scattering, since it is too heavy for the temperatures we are interested in. As the goal of our analysis is to get an estimation of the deviations of the computations using the cBE with respect to the relic density calculated using the standard procedure, i.e. the zero-moment for $f_X(E,T)$, we consider as a benchmark $m_X = 10$ GeV, and $m_\phi \gtrsim 2m_X$.

We implement the sBE and cBEs in Python, and we solve $(Y,y)$ in $x = [10,100]$. In Fig.~\ref{cBEfig}(left) we show the resulting yields using the standard Boltzmann equation (dashed lines) and the coupled Boltzmann equations (eq.~\ref{cBE}), with the black solid line indicating the equilibrium yield. In Fig.~\ref{cBEfig}(right), we show the evolution of the temperature of the dark sector as a function of the photon temperature, with the solid black line indicating the quantity in equilibrium. The first point to highlight here is that for the three values of $m_\phi$ depicted in the plots, both kinetic and chemical decoupling occur nearly at the same time, $x_f \sim 20$, evidencing the \textit{early kinetic decoupling}. Secondly, the benchmark points used in these numerical results enter into the so-called \textit{sub-resonant} regime \cite{Binder:2017rgn}, being those that present the maximum deviations in the relic abundance with respect to the sBE treatment. In this way, this more elaborated analysis in the calculation of the relic abundance shows a significant difference in the relic density calculation in comparison to the standard approach, resulting in relic abundances differences of up to a factor of order one. 

As a detailed and precise analysis over the whole parameter space of the vector DM model is beyond  the scope of our paper, in the rest of our work we consider only the standard procedure in the calculation of the relic abundance, considering the observed DM relic density $\Omega_{\text{DM}} h^2 = 0.1198 \pm 0.0012$ \cite{Planck:2018vyg,Hinshaw:2012aka} as a constraint on the vector relic abundance.

\section{Direct Detection Bounds}
\label{DDB}
\begin{figure}
\hspace{-.45cm}
\begin{minipage}{0.25\textwidth}
\includegraphics[width=\textwidth,angle =-90]{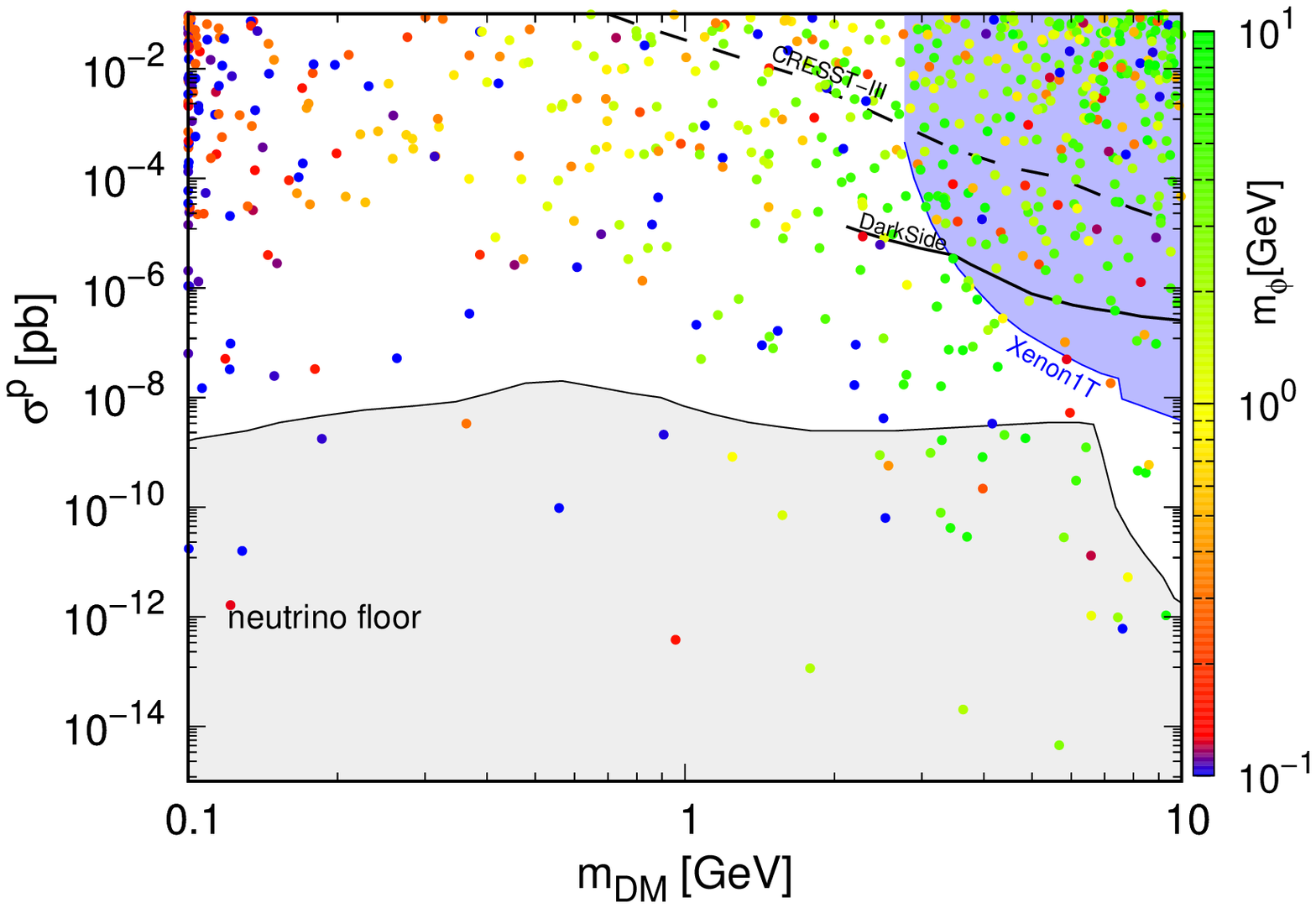}
\end{minipage}
\hspace{1.5cm}
\begin{minipage}{0.25\textwidth}
\includegraphics[width=\textwidth,angle =-90]{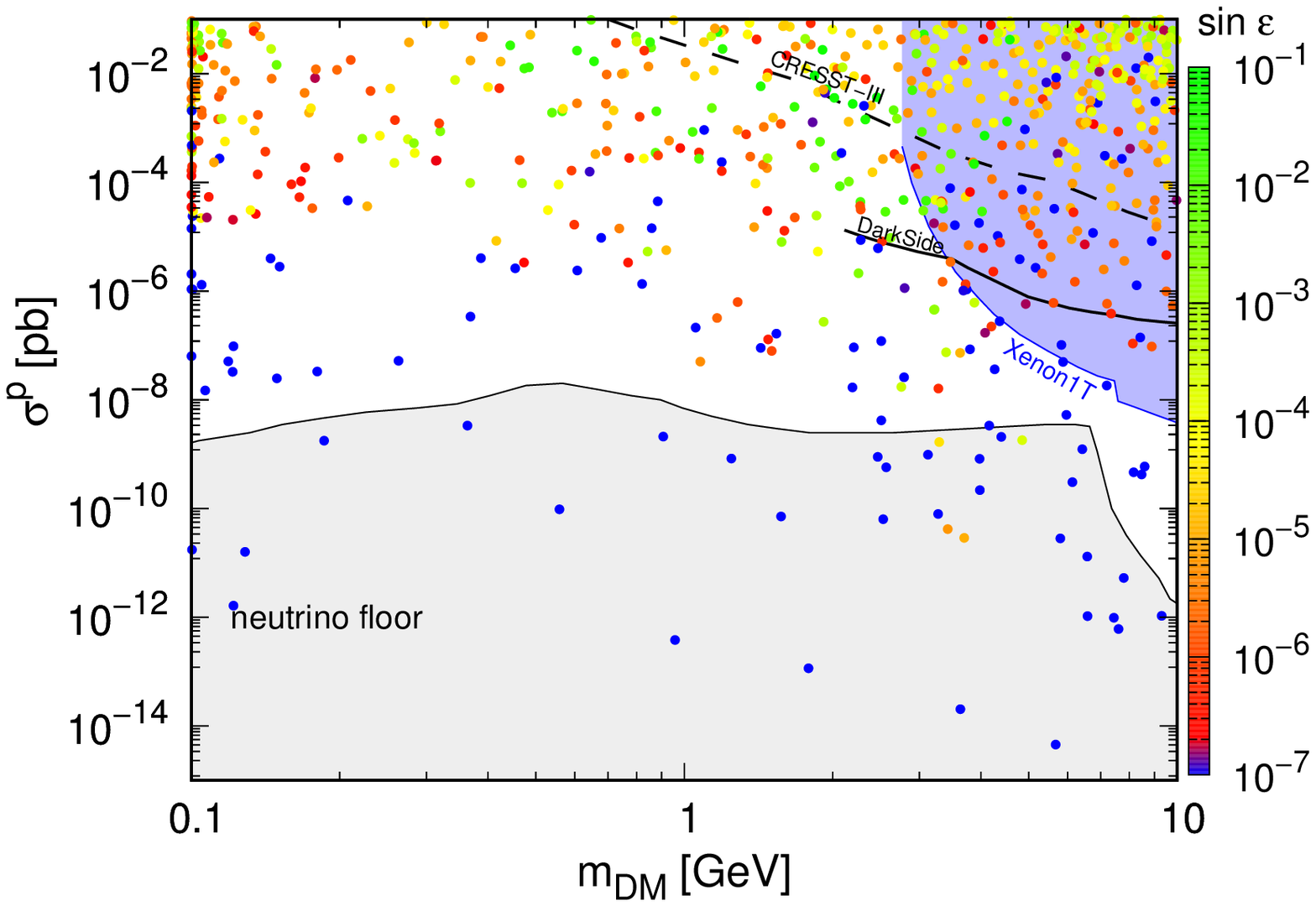}
\end{minipage}
\hspace{1.5cm}
\begin{minipage}{0.25\textwidth}
\includegraphics[width=\textwidth,angle =-90]{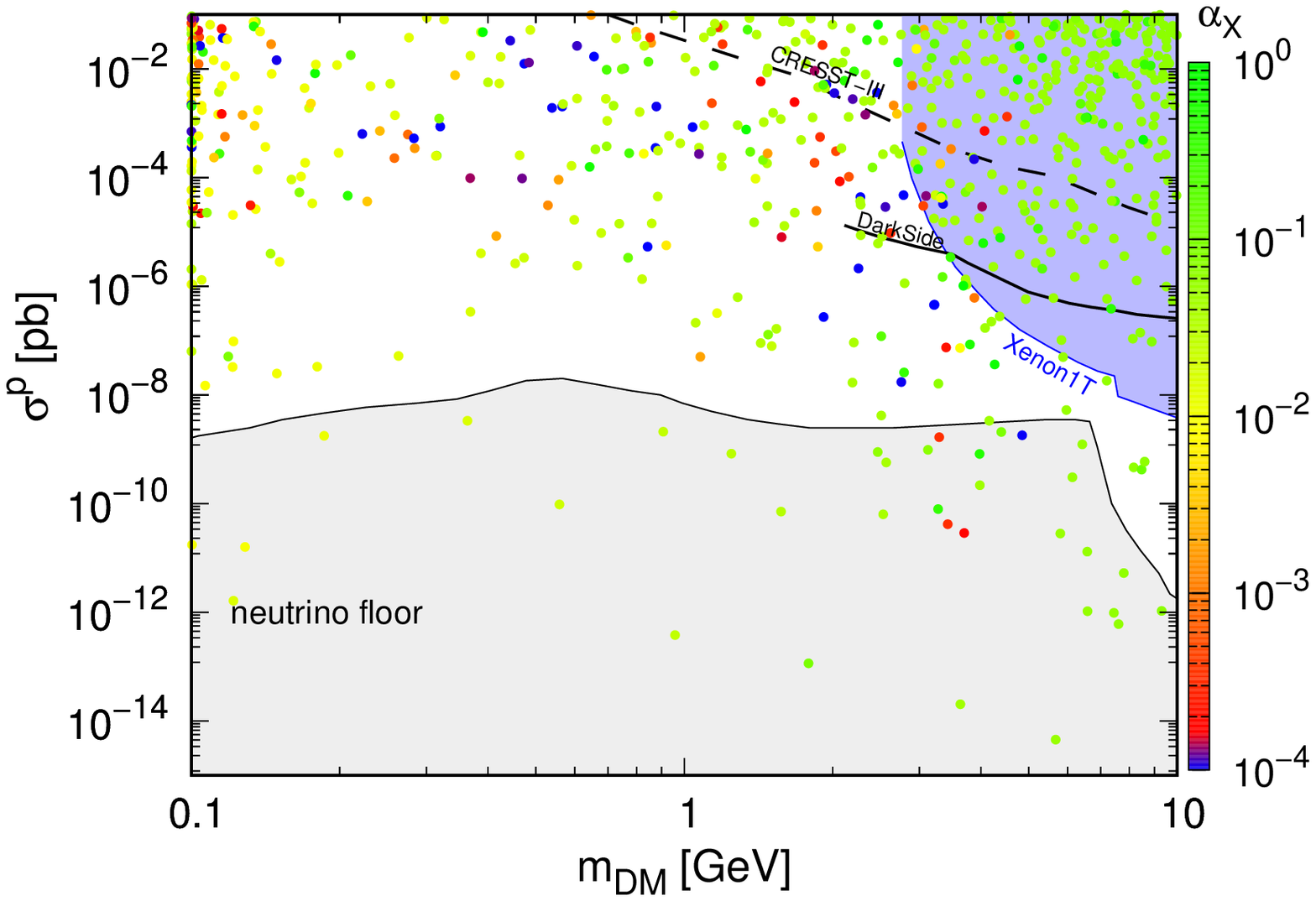}
\end{minipage}
\caption{Viable regions are shown after imposing the observed relic density and invisible Higgs decays bounds. The vertical color spectrum shows the singlet scalar mass ({\it left} panel), the mixing angle ({\it middle} panel), and coupling $\alpha_X$ ({\it right} panel). Direct detection bounds from Xenon1T, DarkSide50 and CRESST-III are applied.} 
\label{DDbounds}
\end{figure}

We consider direct detection bounds on the elastic scattering of DM off the proton from three experiments, Xenon1T \cite{XENON:2019gfn}, DarkSide50-S2 \cite{DarkSide:2018ppu} and CRESST-III \cite{CRESST:2019jnq}.
In the present model the spin-independent (SI) elastic scattering of DM off the normal matter is a $t$-channel process with the SM Higgs or the singlet scalar as the mediator between the DM and the nucleon.
The final formula for the cross sections reads
\begin{equation}
 \sigma^{\text{p}}_{\text{SI}} = \frac{\mu_{Xp}^2 m^2_{p}}{4\pi v_h^2} \alpha_X^2 \sin^2 (2\epsilon) \left(\frac{1}{m_h^2}-\frac{1}{m_\phi^2}\right)^2 f_p^2 \,,
\end{equation}
where $m_p$ is the proton mass, the reduced mass of the proton and DM 
is $\mu_{Xp} = m_X m_p /(m_X+m_p)$, and the scalar form factor for proton is given by 
$f_p  = 2/9 + 7/9 \sum_{q= u,d,s} f^{p}_{q}$, in which $f^{p}_{u} = 0.0153$
$f^{p}_{d} = 0.0191$ and $f^{p}_{s} = 0.0447$ \cite{Belanger:2013oya}.
We probe the parameter space while taking into account bounds from invisible Higgs decay and observed relic density. The ranges of the relevant parameters picked out in our scan 
are the following:  $0.1 < m_{\text{DM}} < 10~$GeV, $0.1 < m_{\phi} < 10~$GeV, $0.001 < \alpha_X < 1$,
and $ 0 < \sin \epsilon < 0.1$. 
The model points in the plane $\sigma^{\text{p}}_{\text{SI}}-m_{\text{DM}}$ of Fig.~\ref{DDbounds} indicate that there are DM masses from 0.1 GeV up 
to 10 GeV in the viable parameter space. The allowed values for the mixing angle in 
the respected regions are smaller than $\sim 10^{-3}$ as shown in the middle panel of Fig.~\ref{DDbounds}. It is also evident from the results in the right panel of Fig.~\ref{DDbounds} that we have for the coupling $\alpha_X \lesssim 0.1$, in the regions respecting the DD bounds. In the final section we will project these same constraints once again. 

\section{CMB Bounds}
\label{CMBbounds}
Light dark matter annihilating into lighter mediators and proceeding in a $s$-wave process can modify the recombination history by injecting energy in form of photons and electrons into the universe \cite{Slatyer:2015jla, Bringmann:2016din}. The anisotropies of the CMB is measured very accurately by Planck \cite{Planck:2018vyg}, and this will put strong indirect constraint on the DM annihilation. The upper limit on the DM annihilation at the recombination is given by 
\begin{equation}\label{cmb}
 \frac{f_{\text{eff}}}{m_{\text{DM}}} \langle \sigma v \rangle_{\text{rec}} < 3.2 \times 10^{-28} \text{cm}^3 \text{s}^{-1} \text{GeV}^{-1}, \,
\end{equation}
where $f_{\text{eff}}$ is an efficiency factor depending on the energy of 
injected electrons and photons. The efficiency factor for models with s-wave annihilation is computed in \cite{Slatyer:2015jla} as a function of DM mass, and 
can be applied to any DM model by weighting their results considering the annihilation product. In our analysis for DM mass below 10 GeV, 
we consider $f_{\text{eff}} = 0.1$ \cite{Padmanabhan:2005es}.

At first sight, it might look impossible to reconcile the thermal relic 
density of about $3\times 10^{-26} \text{cm}^3 \text{s}^{-1}$ to the CMB bound shown in eq.~\ref{cmb}
for DM particles with $s$-wave annihilation and masses below 10 GeV. In fact, if the dark sector is highly decoupled from the SM content, i.e. $\epsilon \ll 1$, then for $m_X > m_\phi$ and $m_X \lesssim 10$ GeV, it occurs that the leading DM annihilation process is into a pair of mediators as shown in Fig.~\ref{FeynAnni}, with $\braket{\sigma_{XX\rightarrow \phi\phi} v} = 3\times 10^{-26}$ cm$^3/$s, in direct conflict with the bound \ref{cmb}.

As discussed in \cite{Griest:1990kh, DAgnolo:2015ujb}, there are two ways to evade the previous CMB bound. The first case is when the DM mass is slightly lighter than 
the mediator mass, then annihilating into heavier states due to the Boltzmann thermal tail, also called forbidden channels, and these processes presenting a suppressed cross section at low velocities. The second case is in which higher partial waves dominate the DM annihilation near a pole. 

\begin{figure}
\center
\includegraphics[width=0.85\textwidth]{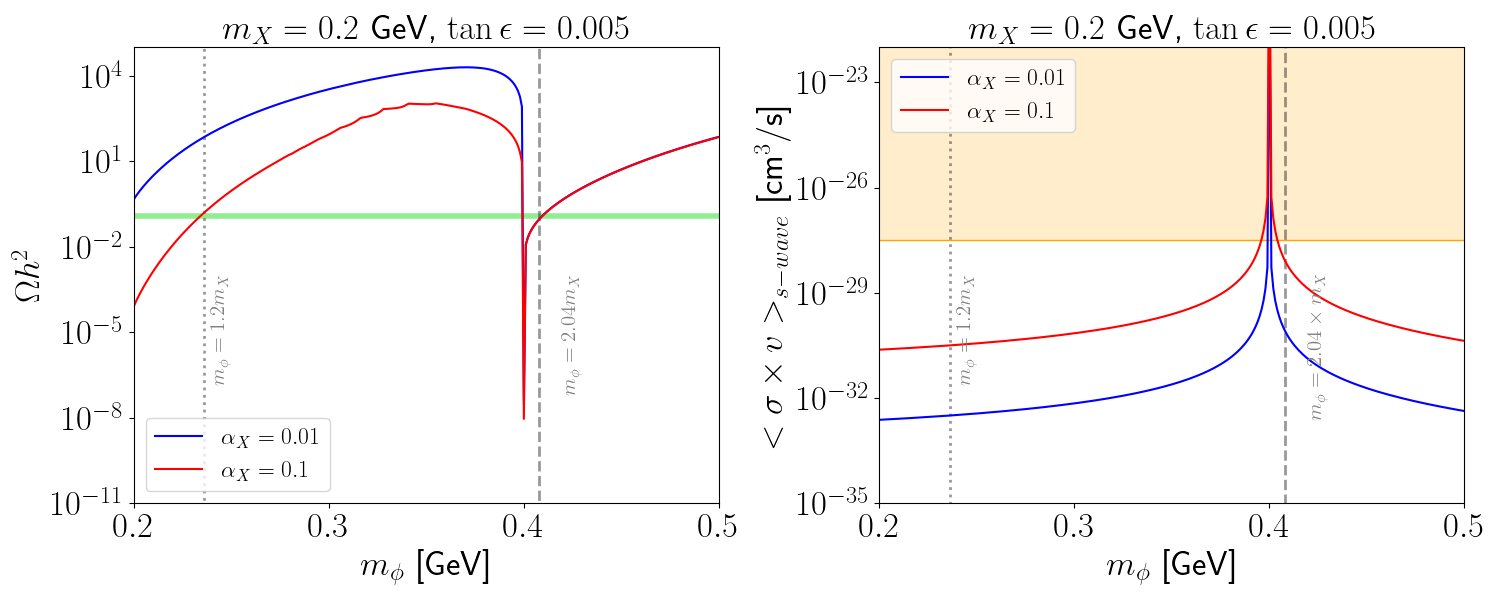}
\caption{In the {\it left plot} we show the relic density as a function of the singlet scalar, and in the {\it right plot} the $s$-wave annihilation cross section times the velocity as a function of the singlet scalar mass at late-times. In both plots we set the DM mass at 0.2 GeV and $\tan \epsilon = 0.005$, with $\alpha_X = 0.1, 0.01$ shown in red and blue, respectively. The observed DM relic density is shown as a horizontal green line at $\Omega h^2 \sim 0.12$ in the left plot, and in the plot in the right we show the CMB bounds as the yellow colored region. The vertical lines in each plot are explained in the text.}
\label{reospnance}
\end{figure}

In order to make more explicit these two ideas, in Fig.~\ref{reospnance} we plot the relic abundance (left plot) and $\langle \sigma v \rangle_{s-wave}$ at late-times (right plot) as a function of the light mediator mass, for $m_X = 0.2$ GeV, $\tan\epsilon = 0.005$ and keeping $m_X < m_{\phi}$. For $\alpha_X = 0.1$ (red line), the correct relic abundance is achieved for two values of $m_\phi$, the first one being slightly heavier than the vector DM mass, $m_\phi/m_{\text{DM}} = 1.20$, with the DM annihilating into the forbidden channels, whereas the other being approximately twice the size of the DM mass, $m_\phi/m_{\text{DM}} = 2.04$, this is, near the light mediator mass pole. As shown in Fig.~\ref{reospnance}(right), these mass values for $m_\phi$ fulfilling the observed relic abundance lay below the CMB bounds (orange region), presenting average cross section sections well below the canonical thermal cross section. Therefore, this type of vectorial DM model for both light DM and light mediators is able to achieve the correct relic abundance fulfilling CMB bounds as long as $m_X < m_\phi$. 

\section{Results}
\label{Secresults}
\begin{figure}
\includegraphics[width=0.57\textwidth]{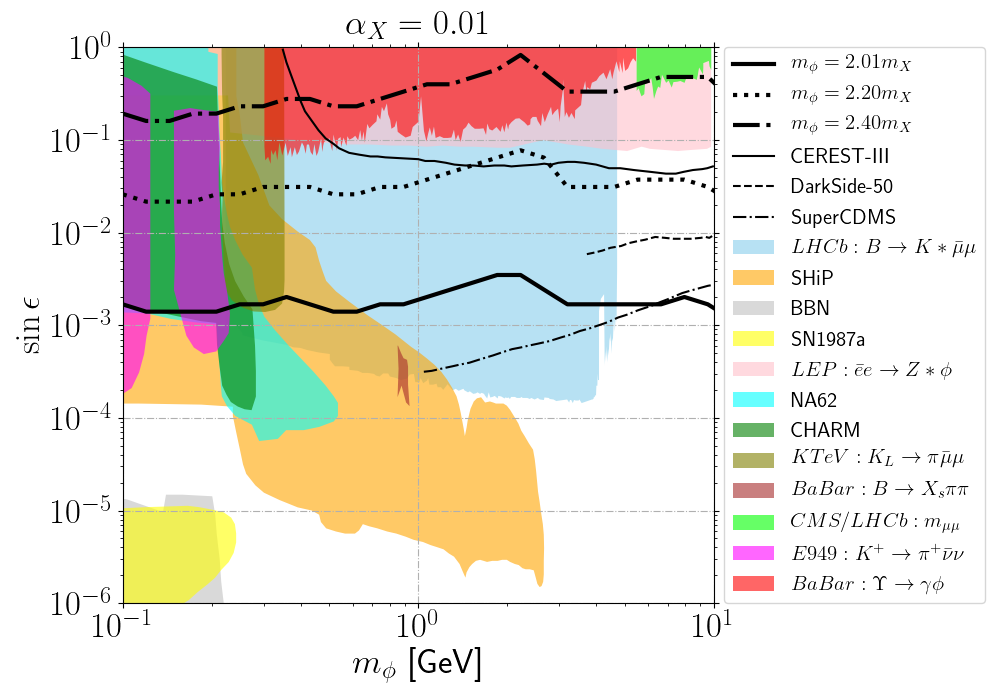}
\includegraphics[width=0.43\textwidth]{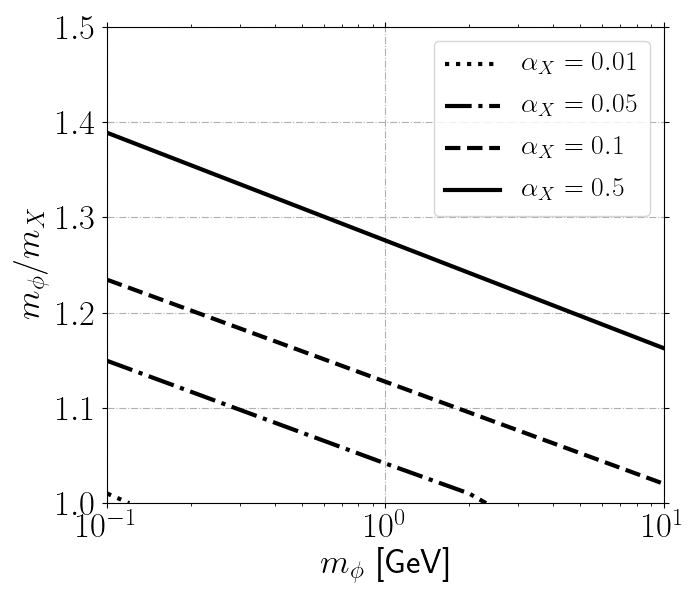}
\caption{(left) Overlaid of predictions in the resonance region (black tick curves) with constraints by experiments and future projections. (right) Points fulfilling the correct relic abundance by the forbidden mechanism, for $\alpha_X$ values specified in the plot.} 
\label{results}
\end{figure}

In here, we summarize the resonance and forbidden viable parameter space of the model under all the relevant constraints. For the former case, we present Fig.~\ref{results} as a representative parameter space viability of the model, considering a few benchmark points of the model fulfilling the correct relic abundance, and the actual constraints. Collider constraints turn out to be too strong, in such a way that most of the viable predictions are excluded by them, even for the case in which the mixing angle $\epsilon$ decreases significantly, e.g. $\mathcal{O}(\epsilon)\sim 10^{-3}$, where we have overlaid the predictions of the model with the tick black curves for different mass shift of the singlet scalar and the DM mass. In particular, provided the mass shift is too small (solid black tick curve), the mixing angle tends to be reduced, but even in this case $B$ meson decay measurements at LHCb are strong. Masses for the new particles above 4 - 5 GeV are not yet constrained by the latter experiment, although future direct detection experiments such as SuperCDMS or DarkSide-50 will be able to test this region. The very low mass region, $\sim 0.1$ GeV is also viable in some discrete range of masses, and experiments such as SHiP or CHARM could explore these parameter space. Thermalization condition is not competitive to the present constraints shown in Fig.~\ref{results}(left), since they impose $\epsilon \lesssim \mathcal{O}(10^{-6})$. The details of our thermalization results can be seen in App.~\ref{ApenC}.

The forbidden DM relic abundance is dependent only on three parameters: $m_X, m_{\phi}$ and $\alpha_X$. In order to exemplify the relation of them fulfilling the observed relic abundance, in Fig.~\ref{results}(right) we show some results obtained with MicrOMEGAS code in the mass plane for different values of $\alpha_X$. This behavior is understood by the interplay between the parameters such that the relic density is kept constant \cite{DAgnolo:2015ujb}. Note that too small $\alpha_X$ requires a too high degeneracy between the new components of the new sector, and a very limited range of masses. This is in agreement with the results of scalar version of this model \cite{Hara:2021lrj}. Since the relic abundance produced by the forbidden channel does not require any particular value of $\epsilon$, one can arbitrarily take too small values of it in such a way to avoid all the constraints that applied to the resonance case, then entering into the still allowed regions of Fig.~\ref{results}(left). In this case, BBN and SN1987a are important constraints to take into account. Finally, although not obligatory, thermalization criteria could set bounds on the mixing angle too, but this analysis is beyond the scope of our work.

\section{Conclusions}
\label{con}
In this work, we reconsidered the Higgs portal $U(1)_X$ vector dark matter 
model regarding the low mass of the DM candidate and the singlet scalar. 
This region of the parameter space with $m_{\text{DM}}$ below 10 GeV 
is not fully investigated in earlier works. In this region of the parameter space, even when a vast parameter space is able to fulfill the measured relic abundance, cosmic microwave background bounds set particularly strong constraint, obligating the model to adjust its relic only in two places via thermal freeze-out: resonance DM annihilation and forbidden channel annihilations. The invisible Higgs decay bounds becomes strong and to respect the 
corresponding limits, quite small mixing angle is demanded. Other collider searches 
in $B$ and $K$ meson are considered along with the limits from beam-dump experiments. Since the singlet scalar can decay to the SM leptons we ensured that 
it will not spoil the BBN.  
Our main results given in Fig.~\ref{results} concern the viable regions respecting all the theoretical and experimental constraints. 

We found that when the DM annihilations are near the pole resonance, various parameter space regions get excluded, and in the future direct detection constraints will be able to probe most of the region of DM masses $\mathcal{O}(1)$ GeV, whereas lower masses will be tested by experiments such as SHiP and CHARM. On the other hand, regarding the forbidden dark matter production, it shows that as it is independent of the mixing angle $\epsilon$, it is less constrained than the parameter space originated by the resonance production, therefore being more challenging to test regions with such small $\epsilon$.

Finally, as a matter of precision, we have checked that, as light DM in this context requires low mixing angles $\epsilon$, the vector DM in the resonance annihilation region presents early loss of kinetic equilibrium, changing the standard calculation of the relic abundance up to one order of magnitude in the relic density. A full analysis considering this level of precision and including the  effects of this early kinetic decoupling on the forbidden mechanism is left for a future work.
\section{Acknowledgments}
B.D.S has been founded by ANID (ex CONICYT) Grant No. 74200120.

\appendix
\section{DM Annihilation Cross Sections}
\label{ApenA}
Vector DM annihilation cross sections times the relative velocity are provided in this section making use of CalcHEP code. The first formula is the annihilation of DM to a fermion pair being a $s$-channel process:
\begin{equation}\label{XXff}
\small
 \sigma v_{rel} (X X \to f^+ f^-) = \frac{N_p N_c m_f^2 \alpha_{X}^2 (s - 4m_f^2)(12m_X^4 - 4m_X^2 s + s^2)\sin^2(2\epsilon)}{48\pi m_X^4v_h^2}\frac{1}{(s-m^2_\phi)^2 + m_\phi^2 \Gamma_\phi^2} \,,
  \end{equation}
where $N_p = 9$ is the number of initial vector polarizations, $N_c$ is the number of colors in case of quarks in the final state, and $\Gamma_\phi$ is the total width of the light mediator $\phi$. As we focus in the low mass regime, i.e. $\sqrt{s} \ll m_h$, we neglect the Higgs contribution for the annihilation cross section.

The next formula belongs to the DM annihilation to a pair of singlet scalars in $s$-, $t$-
and $u$-channel: 
\begin{equation}
\begin{aligned}
 \sigma v_{rel} (X X \to s s) & = \frac{\sqrt{1-4m^2_s/s}}{32\pi^2s}
\int d\Omega \Big[ \frac{8}{9} \cos^4(\epsilon) \alpha_{X}^4 + \frac{64}{9} \cos^4(\epsilon) w^4 \alpha_{X}^8 \Big(\frac{1}{t-m_X^2} + \frac{1}{u-m_X^2}\Big)^2
  \\ &
    + 8 w^2 \alpha_X^4 \Big(\frac{c_1 \cos(\epsilon)}{s-m_\phi^2} - 
   \frac{c_2 \sin(\epsilon)}{s-m_h^2} \Big)^2
  -\frac{64}{9} \cos^4(\epsilon) w^2 \alpha_{X}^6 \Big(\frac{1}{t-m_X^2}+\frac{1}{u-m_X^2}\Big)  
  \\&
 - \frac{16c_1 \cos^3(\epsilon) w \alpha_X^4}{3(s-m_X^2)}
 + \frac{16c_2 \cos^2(\epsilon) \sin(\epsilon) w \alpha_X^4}{9(s-m_h^2)}
\\&
 +\frac{64c_1 \cos^3(\epsilon)w^3 \alpha_X^6}{3(s-m_\phi^2)}\Big(\frac{1}{t-m_X^2} 
  +\frac{1}{u-m_X^2}\Big)
 \\&
 -\frac{64c_2 \cos^2(\epsilon) \sin(\epsilon) w^3 \alpha_X^6}{9(s-m_h^2)}\Big(\frac{1}{t-m_X^2} + \frac{1}{u-m_X^2}\Big)
 \Big] \,,
\end{aligned}
\end{equation}
where, the couplings $c_1$ is
\begin{equation}
 c_1 = 2 v_h \lambda_H \sin^3(\epsilon)  + w \lambda_{HS} \cos(\epsilon) \sin^2(\epsilon)   
 +2 w \lambda_S \cos^3(\epsilon) \,,
 \end{equation}
and the couplings $c_2$ is
\begin{equation}
 \begin{aligned}
  c_2 &= v_h \lambda_{HS} \cos(\epsilon) + 2 w \lambda_{HS} \sin(\epsilon)
 -3 w  \lambda_{HS} \sin^3(\epsilon) + 6 v_h \lambda_{H} \cos(\epsilon) \sin^2(\epsilon)
 \\&
 - 3 v_h  \lambda_{HS} \cos(\epsilon) \sin^2(\epsilon) 
 - 6 w  \lambda_{S} \cos^2(\epsilon) \sin(\epsilon) \,.
 \end{aligned}
\end{equation}

\section{Temperature-weighted thermal average}
\label{appb}
The second annihilation cross section relevant for the early kinetic decoupling is given by \cite{Binder:2017rgn}
\begin{eqnarray}\label{sigv2a}
 \braket{\sigma v}_2 \equiv \Braket{\sigma v \frac{\bold{p}^2}{3E}} &\equiv & \frac{g_X^2}{T n_{X,e}^2}\int \frac{d^3p}{(2\pi)^3}\frac{d^3\tilde{p}}{(2\pi)^3} (\sigma v)\frac{p^2}{3E}f_{X,e}(\bold{p})f_{X,e}(\tilde{\bold{p}}) ,
\end{eqnarray}
with $g_X$ the internal degrees of freedom of the DM particle, and $f_{X,e}(\bold{p})$ corresponding to the equilibrium density $\sim\exp(-E/T)$ (equivalently for $f_{X,e}(\tilde{\bold{p}})$). We use the result obtained in \cite{Yang:2019bvg}:
\begin{eqnarray}\label{sigv2}
 \braket{\sigma v}_2 &\approx& \frac{1}{48m_X^4 K_2^2(m_X/T)} \int_{4m_X^2}^\infty ds (\sigma v_{lab})\sqrt{s - 4m_X^2} \\
 &\times &\left[(s + 2m_X^2)K_1\left(\frac{\sqrt{s}}{T}\right) + \left(\frac{(s - 4m_X^2)\sqrt{s}}{2T} + \frac{4T(s + 2m_X^2)}{\sqrt{s}}\right)K_2\left(\frac{\sqrt{s}}{T}\right)\right] 
\end{eqnarray}

\section{Vector DM thermalization}\label{ApenC}
In this section we explore the resulting constraint on the parameter space after imposing a thermalization condition. This is, we demand that at the freeze-out epoch of the DM, $T_f = m_X/x_f \sim m_\phi/x_f$ with $x_f \sim 15 - 25$, the rate of interactions involving the mediator $\phi$ must be bigger than the expansion of the universe\footnote{Two comments should be made here. First, as the interactions between the mediator $\phi$ and the vector DM $X_\mu$ are mediated by $\alpha_X\cos\epsilon$, and considering $\alpha_X \gtrsim 0.01$, it is sufficient to explore the thermalization conditions on the mediator, then automatically interactions $X_\mu - \phi$ will bring the former into thermal equilibrium. Secondly, the evaluation of the rates at the freeze-out temperature is based on the assumption that the rate of interactions at higher temperatures goes smaller than $T^2$.}. In particular, we focus on the following processes:
\begin{enumerate}
 \item Decays: $\phi\rightarrow$ SM,
 \item Scattering $\phi + f \rightarrow \phi + \gamma$.
\end{enumerate}
Here, $f$ represents a SM fermion. In mathematical terms, the thermalization condition is given by 
\begin{eqnarray}\label{thermalcond}
 H(T_f) < \Gamma_{\phi \rightarrow \text{SM}}(T_f) + \Gamma_{\phi + SM \rightarrow \phi + \text{SM'}}(T_f).
\end{eqnarray}
The two-body decays are split in two parts: 
\begin{eqnarray}
 \Gamma_{\phi \rightarrow \text{SM}}(T) \equiv \Gamma_{\phi \rightarrow \gamma\gamma}(T) + \sum_\text{fermions} \Gamma_{\phi \rightarrow f + \bar{f}}(T),
\end{eqnarray}
with 
\begin{eqnarray}
 \Gamma(\phi\rightarrow \gamma\gamma) = \frac{\sin^2\epsilon \alpha^2 m_{\phi}^3}{256\pi^3 v_h^2}\vert f(z_e) + \frac{7}{3}|^2,
\end{eqnarray}
where $\alpha$ is the fine-structure constant, $f(\tau) = 2\tau \left[1 + (1 - \tau)\arctan^2(1/\sqrt{\tau - 1})\right]$, and $z_e = 4m_e^2/m_{\phi}^2$, whereas the decay into fermions is given in \ref{phiff}\footnote{In this analysis we assume that $m_\phi \gtrsim m_X$ in such a way that the decay $\phi \rightarrow XX$ is not kinematically allowed.}. On the other hand, the cross section for the scattering was calculated with CalcHEP, and as it is more involved in terms of algebra, so we do not include it here. Notice that the rate of scattering interactions is given by $\Gamma_{scatt} = \sum_f n_f \braket{\sigma v_{rel}}$, where the relevant interactions consider that the fermion in the plasma is still relativistic, i.e. $n_f = \zeta(3)g T^3/\pi^2$, with $g$ the internal degree of freedom of the mediator.
\begin{figure}[h!]
\centering
\includegraphics[width=0.5\textwidth]{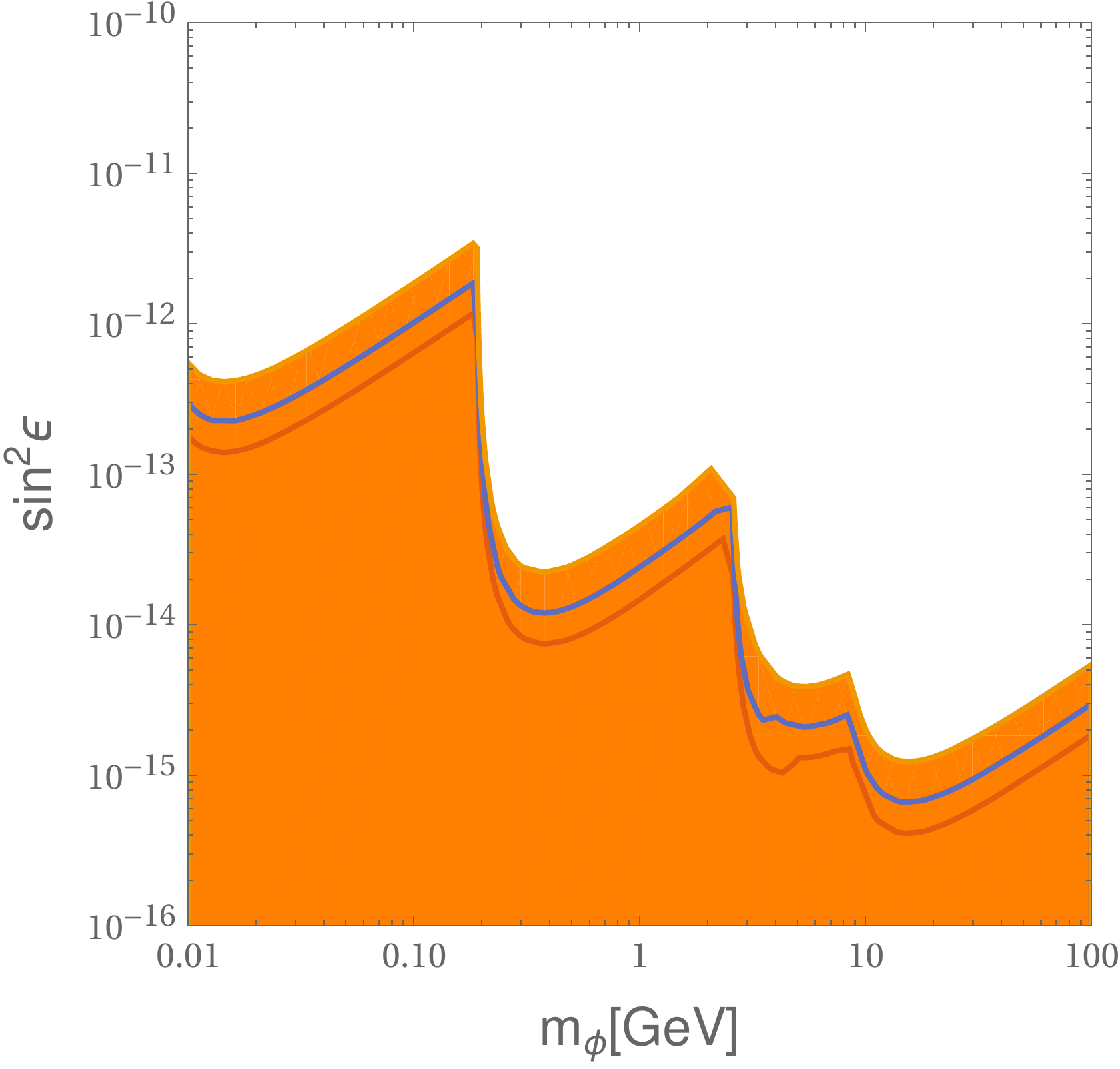}
\caption{Thermalization condition given in eq.~\ref{thermalcond}. The yellow, blue and red curves correspond to $x_f = 15, 20$ and 25, respectively.}
\label{thermalplot}
\end{figure}

The resulting exclusion in the parameter space is shown in Fig.~\ref{thermalplot}, where all the colored region does not fulfill the condition of eq.~\ref{thermalcond}. In the figure we have included the exclusion for $x_f = 15, 20$ and $25$, showing a similar pattern for the three cases, although for a determined value of $m_\phi$, the allowed mixing angle $\epsilon$ can vary up to one order of magnitude. This result will be useful as a floor for $\epsilon$ when we include the full parameter space with all the constraints. Similar to the results in \cite{Hara:2021lrj}, we have found that scattering collisions are subleading in comparison to the decay process by several orders of magnitude, although our constraint is stronger in comparison to what is found in \cite{Hara:2021lrj}. In contrast to \cite{Hara:2021lrj}, in our work we have considered most of the relevant thresholds resulting from the two-body decay of $\phi$ into SM particles, including the photons. Pions were not included.

\bibliography{ref}

\bibliographystyle{utphys}

\end{document}